\documentstyle[12pt,epsf]{article}
\begin{document}
\newcounter{myfn}[page]
\renewcommand{\thefootnote}{\fnsymbol{footnote}}
\newcommand{\myfootnote}[1]{\setcounter{footnote}{\value{footnote}}%
\footnote{#1}\stepcounter{footnote}}
\newcounter{saveeqn}
\newcommand{\add}{\addtocounter{equation}{1}}
\newcommand{\alpheqn}{\setcounter{saveeqn}{\value{equation}}%
\setcounter{equation}{0}%
\renewcommand{\theequation}{\mbox{\thesection.\arabic{saveeqn}{\alph{equation}}}}}
\newcommand{\reseteqn}{\setcounter{equation}{\value{saveeqn}}%
\renewcommand{\theequation}{\thesection.\arabic{equation}}}
\newenvironment{nedalph}{\add\alpheqn\begin{eqnarray}}{\end{eqnarray}\reseteqn}
\renewcommand{\theequation}{\thesection.\arabic{equation}}
\newsavebox{\PSLASH}
\sbox{\PSLASH}{$p$\hspace{-2.1mm}/}
\newcommand{\PS}{\usebox{\PSLASH}}

\thispagestyle{empty}
\begin{flushright}
HD--THEP--96--39
\par
hep--lat/9610001
\end{flushright}
\vspace{1cm}
\begin{center}
{\large\bf{Continuum Behaviour of Lattice QED, Discretized with}}\\
\vspace{.3cm}
{\large\bf{One--Sided Lattice Differences, in One--Loop Order}}\\
\vspace{0.5cm}
N\'eda Sadooghi\footnote{\normalsize{e-mail: n.sadooghi@thphys.uni-heidelberg.de}} and Heinz J. Rothe\footnote{\normalsize{e-mail: h.rothe@thphys.uni-heidelberg.de}}\\
\vspace{0.5cm}
{\sl{ Institut f\"ur Theoretische Physik}}\\
{\sl{Universit\"at Heidelberg}}\\
{\sl{Philosophenweg 16, 69120 Heidelberg, Germany}}\\
\end{center}
\vspace{1cm}
\begin{center}
{\bf {Abstract}}
\end{center}
\begin{quote}
A lattice action for QED is considered, where the derivatives in the Dirac operator are replaced by one--sided lattice differences. A systematic expansion in the lattice spacing of the one--loop contribution to  the fermion self energy, vacuum polarization tensor and vertex function is carried out for an arbitrary choice of one--sided lattice differences. It is shown that only the vacuum polarization tensor possesses the correct continuum limit, while the fermion self energy and vertex function receive  non--covariant contri\-butions. A lattice action, discretized with a fixed choice of one--sided lattice differences therefore does not define a renormalizable field theory. 
The non-covariant contributions can however be eliminated by averaging the expressions over all possible choices of one--sided lattice differences. 
\end{quote}

\par
PACS number(s): 11.15.-q; 11.15.Ha

\par
Keywords: Lattice fermions, Lattice perturbation theory.

\newpage

\section{Introduction}
\setcounter{footnote}{0}
\hspace{1cm}
As is well known, the naive discretization of the fermionic action which preserves the antihermiticity of the derivatives in the continuum, leads to the notorious fermion doubling problem. One of the most popular proposals for eliminating the fermion--doubling problem consists in working with Wilson fermions \cite{wilson}, whereby the action is modified by  a so--called  irrelevant term which breaks explicitely the chiral symmetry. This is consistent with the Nielson--Ninomiya theorem which states that for an action respecting the usual translational invariance, locality and hermiticity requirements,  the fermion doubling problem can only be eliminated by breaking chiral symmetry \cite{nn}. Another way to circumvent the above theorem is to give up reflection positivity. Thus  replacing the derivatives in the continuum by one--sided lattice differences ensures that high momentum excitations arising from the corners of the Brillouine zone do not contribute in the continuum limit. In fact the functional formalism of Berezin \cite{berezin} dictates that the temporal derivative in the Dirac operator should be discretized using the one--sided lattice differences. By symmetry one would then be led to a similar discretization of the spatial derivatives. In this case the Hamiltonian would no longer be hermitian for finite lattice spacing, which does, however, not necessary imply that correlation functions do not possess the correct continuum limit. Actions involving one--sided lattice differences have been considered  in \cite{jacobs, stam, De}, where by a suitable averaging procedure, reflection symmetry is effectively restored a posteriori. 
\par
In this paper we will study in detail the behaviour of the fermion self energy, vacuum polarization tensor and vertex function  for small lattice spacing in one--loop order, for a lattice action 
discretized with one-sided differences. This analysis is carried out using the small--$a$--expansion scheme discussed in \cite{ww}. Such a systematic study has not been  presented in the literature, which is surprising in view of the importance of the problem.  It is shown  that while the vacuum polarization has the expected continuum form for small lattice spacing, the fermion self energy and vertex function include non--covariant contributions which can only be eliminated by averaging  the expressions over all possible choices of one--sided lattice differences. These computations are carried out in sections \ref{sec3} and \ref{sec4}.
Since the small--$a$--expansion scheme of ref. \cite{ww} has not been widely used in the literature, we shall discuss it in detail in the following section. This section also serves to define our notations used in sections \ref{sec3} and \ref{sec4}. 

 
\section{Small--$a$--expansion}\label{sec2}
 \hspace{1cm}
In this section, we discuss in detail the small--$a$--expansion scheme of ref. \cite{ww} 
following closely the proof given in that reference. 
\par
A lattice regulated one--loop Feynman integral in $D$ dimension has the form:
\begin{eqnarray}\label{A1}
\sigma\left(p,m,a,D\right)&=&\int\limits_{-\frac{\pi}{a}}^{+\frac{\pi}{a}}\frac{d^{D}p'}{\left(2\pi\right)^{D}}\ 
H\left(p', p, m,  a, D \right),
\end{eqnarray}
where $p$ stands for the set of external momenta, $p'$ is the loop momentum and $a$ the lattice spacing. The particle masses are denoted collectively by $m$. An important feature of the lattice regulated Feynman integrals is that the lattice spacing appears in the integrand as well as in the integration limits. This integrand is a periodic function of $p$ and $p'$ and possesses a continuum limit. Furthermore $H\left(p', p, m,  a, D \right)$ has the following homogeinity property, 
\begin{eqnarray}\label{A2}
H\left(p', p, m,  a, D \right)&=&a^{d_{H}} H\left(ap', ap, am,a=1, D \right),
\end{eqnarray}
where $d_{H}$ is the inverse--mass dimension of the integrand. In the 
{\it naive} continuum limit the integral (\ref{A1}) reduces to the usual, in general divergent, continuum Feynman integral. Since $H\left(p', p, m,  a, D \right)$ is an  analytic function of $a$ in the neighborhood of $a=0$, we can expand it into a Taylor series. Consider the expansion up to ${\cal{O}}\left(a^{J}\right)$,
\begin{eqnarray}\label{A3}
H\left(p', p,m,a,D\right)&=&\sum\limits_{j=0}^{J}a^{j}H_{j}\left(p', p,m,D\right)+R_{J}\left(p', p,m,a,D\right).
\end{eqnarray}
The Taylor coefficients $H_{j}\left(p', p,m,D\right)$, which have a structure resembling that of the integrands of continuum Feynman integrals,  possess the following homogeinity property:
\begin{eqnarray}\label{A4}
H_{j}\left(ap', ap, am, D\right)&=&a^{j-d_{H}}H_{j}\left(p', p, m,D\right).
\end{eqnarray}
From (\ref{A2}) and (\ref{A4}) it further follows for the remainder $R_{J}$ that:
\begin{eqnarray}\label{A5}
R_{J}\left(p', p,m,a,D\right) &=&a^{d_{H}}R_{J}\left(ap', ap,am,a=1,D\right).
\end{eqnarray}
Introducing (\ref{A3}) into (\ref{A1}) we have that:
\begin{nedalph}\label{A6a}
\sigma\left(p,m,a,D\right)&=&
\sum\limits_{j=0}^{J}\sigma_{j} \left(p,m,a,D\right) +\int\limits_{-\frac{\pi}{a}}^{+\frac{\pi}{a}}\frac{d^{D}p'}{\left(2\pi\right)^{D}}\ R_{J}\left(p', p,m,a,D\right),
\end{eqnarray}
where
\begin{eqnarray}\label{A6b}
\sigma_{j} \left(p,m,a,D\right) &=&a^{j}\int\limits_{-\frac{\pi}{a}}^{+\frac{\pi}{a}}\frac{d^{D}p'}{\left(2\pi\right)^{D}}\ H_{j}\left(p',p,m,D\right).
\end{nedalph}
Consider first (\ref{A6b}). This expression can be decomposed as follows:
\begin{eqnarray}\label{A7}
\sigma_{j}\left( p,m,a,D\right) &=& a^{j} \bigg[\int\limits_{-\infty}^{+\infty}\frac{d^{D}p'}{\left(2\pi\right)^{D}}\  H_{j}\left(p', p,m,D\right)-
\int\limits_{|p'|>\frac{\pi}{a}} 
\frac{d^{D}p'}{\left(2\pi\right)^{D}}\  H_{j}\left(p', p,m,D\right)\bigg].\nonumber\\
\end{eqnarray}
For $j=0$, the first term corresponds to the naive continuum limit of (\ref{A1}). For $D=4$ this integral in general diverges. Hence from now on it will be understood that all integrals are regulated using the dimensional regularization scheme\footnote{\normalsize{Dimensional regularization of lattice Feynman integrals have also been considered by Kawai et al. \cite{kawai}}}. Notice that  for $a\neq 0$ and $m\neq 0$ the original lattice Feynman integral (\ref{A1}) will be  well defined.  Hence any divergencies introduced by this decomposition have to cancel eventually.  
\par
Making use of the homogeneity property (\ref{A4}), and introducing the dimensionless integration variable $\hat{p}'=ap'$, expression (\ref{A7}) takes the form:
\begin{eqnarray*}
\lefteqn{\sigma_{j}\left( p,m,a,D\right) =}\nonumber\\
&&= a^{j} \int\limits_{-\infty}^{+\infty}\frac{d^{D}p'}{\left(2\pi\right)^{D}}\  H_{j}\left(p', p,m,D\right)-a^{d_{H}-D}
\int\limits_{|\hat{p}'|>\pi} 
\frac{d^{D}\hat{p}'}{\left(2\pi\right)^{D}}\  H_{j}\left(\hat{p}', ap,am,D\right).
\end{eqnarray*}
We next expand $ H_{j}\left(\hat{p}', ap,am,D\right)$ in the lattice spacing $a$  up to ${\cal{O}}\left(a^{J-\left(d_{H}-4\right)}\right)$. Then
\begin{eqnarray}\label{A8}
\lefteqn{\sigma_{j}\left( p,m,a,D\right)=}\nonumber\\
 &=& a^{j} \int\limits_{-\infty}^{+\infty}\frac{d^{D}p'}{\left(2\pi\right)^{D}}\  H_{j}\left(p', p,m,D\right)-a^{d_{H}-D}
\int\limits_{|\hat{p}'|>\pi} 
\frac{d^{D}\hat{p}'}{\left(2\pi\right)^{D}}\  T_{J-\left(d_{H}-4\right)}H_{j}\left(\hat{p}', ap,am,D\right)\nonumber\\
&&+{\cal{O}}\left(a^{J+\left(4-D\right)+1}\right),
\end{eqnarray}
where $T_{k}f\left(a\right)$ stands for the Taylor expansion of $f\left(a\right)$ around $a=0$ up to order $k$.   The expression (\ref{A8}) can be trivially rewritten as follows
\begin{eqnarray}\label{A9}
\lefteqn{\sigma_{j}\left( p,m,a,D\right)=}\nonumber\\
 &=& a^{j} \int\limits_{-\infty}^{+\infty}\frac{d^{D}p'}{\left(2\pi\right)^{D}}\  H_{j}\left(p', p,m,D\right)+a^{d_{H}-D}
\int\limits_{-\pi}^{+\pi} 
\frac{d^{D}\hat{p}'}{\left(2\pi\right)^{D}}\  T_{J-\left(d_{H}-4\right)}H_{j}\left(\hat{p}', ap,am,D\right)\nonumber\\
&&-a^{d_{H}-D}
\int\limits_{-\infty}^{+\infty} 
\frac{d^{D}\hat{p}'}{\left(2\pi\right)^{D}}\  T_{J-\left(d_{H}-4\right)}H_{j}\left(\hat{p}', ap,am,D\right) +{\cal{O}}\left(a^{J+\left(4-D\right)+1}\right). 
\end{eqnarray}
The $\hat{p}' $--dependence of the coefficients in the Taylor expansion of $T_{J-\left(d_{H}-4\right)}H_{j}$ have the form 
$\hat{p}' _{\mu_{1}} \cdots \hat{p}' _{\mu_{l}}/\left(\hat{p}^{'2}\right)^{n} $.
Hence the last integral vanishes in the dimensional regularization scheme. Introducing (\ref{A9}) into (\ref{A6a}), we therefore have that:
\begin{eqnarray}\label{A10}
\lefteqn{\sigma\left( p,m,a,D\right)=}\nonumber\\
 &&=\sum\limits_{j=0}^{J}\bigg\{a^{j}\int\limits_{-\infty}^{+\infty}\frac{d^{D}p'}{\left(2\pi\right)^{D}}H_{j}\left(p', p,m,D\right)+a^{d_{H}-D}
\int\limits_{-\pi}^{+\pi} 
\frac{d^{D}\hat{p}'}{\left(2\pi\right)^{D}}T_{J-\left(d_{H}-4\right)}H_{j}\left(\hat{p}', ap,am,D\right)\bigg\}\nonumber\\
&&+a^{d_{H}-D}
\int\limits_{-\pi}^{+\pi} 
\frac{d^{D}\hat{p}'}{\left(2\pi\right)^{D}}\  R_{J}\left(\hat{p}', ap,am,a=1,D\right) +{\cal{O}}\left(a^{J+\left(4-D\right)+1}\right), 
\end{eqnarray}
where we have made use of the homogeneity property (\ref{A5}),  and have introduced the dimensionless loop momentum $\hat{p}' =ap'$ in the last integral in (\ref{A6a}). From (\ref{A2}), (\ref{A4}) and (\ref{A5}) it follows that:
\begin{eqnarray*}
\sum\limits_{j=0}^{J}H_{j}\left(\hat{p}', ap,am,D\right)&=&H\left(\hat{p}', ap,am,a=1,D\right)-R_{J}\left(\hat{p}', ap,am,a=1,D\right).
\end{eqnarray*}
Introducing this expression into (\ref{A10}), we therefore have that:
\begin{eqnarray*}
\lefteqn{\sigma\left( p,m,a,D\right)= \sum\limits_{j=0}^{J}a^{j} \int\limits_{-\infty}^{+\infty}\frac{d^{D}p'}{\left(2\pi\right)^{D}}\  H_{j}\left(p', p,m,D\right)}\nonumber\\
&&+a^{4-D}\bigg[a^{d_{H}-4}
\int\limits_{-\pi}^{+\pi} 
\frac{d^{D}\hat{p}'}{\left(2\pi\right)^{D}}\  T_{J-\left(d_{H}-4\right)}H\left(\hat{p}', ap,am,a=1, D\right)\bigg]+{\cal{O}}\left(a^{J+\left(4-D\right)+1}\right).  
\end{eqnarray*}
Hence the small--$a$--expansion scheme can be summarized as follows:
\begin{nedalph}\label{A11a}
\sigma\left(p,m,a,D=4\right)&=&\lim\limits_{D\to 4}\bigg[
\sigma^{\left(\infty\right)}\left(p,m,a,D\right)+a^{4-D}   \tilde{\sigma}\left(p,m,a,D\right)
\bigg],
\end{eqnarray}
where
\begin{eqnarray*}
\sigma^{\left(\infty\right)}\left(p,m,a,D\right) &\equiv&\sum\limits_{j=0}^{J} a^{j}
\sigma^{\left(\infty\right)/\left(j\right)}\left(p,m,D\right) + {\cal{O}}\left(a^{J+1}\right),
\end{eqnarray*}
with
\begin{eqnarray}\label{A11b}
\sigma^{\left(\infty\right)/\left(j\right)}\left(p,m,D\right)&=&\frac{1}{j!}\int\limits_{-\infty}^{+\infty}
\frac{d^{D}p'}{\left(2\pi\right)^{D}}\  \ \left(
\frac{\partial^{j}H\left(p',p,m,a,D\right)}{\partial a^{j}}
\right)_{a=0},
\end{eqnarray}
and
\begin{eqnarray*}
\tilde{\sigma}\left(p,m,a,D\right)&\equiv& \sum\limits_{j=d_{H}-4}^{J} a^{j}
\tilde{\sigma}^{\left(j\right)}\left(p,m,D\right) + {\cal{O}}\left(a^{J+1}\right),
\end{eqnarray*}
where
\begin{eqnarray}\label{A11c}
\tilde{\sigma}^{\left(j\right)}\left(p,m,D\right)&=&\frac{1}{\left(j-d_{H}+4\right)!} \int\limits_{-\pi}^{+\pi}
\frac{d^{D}\hat{p}'}{\left(2\pi\right)^{D}}
 \ \left(
\frac{\partial^{j-d_{H}+4}H\left(\hat{p}',ap,am,a=1,D\right)}{\partial a^{j-d_{H}+4}}
\right)_{a=0}.\nonumber\\
\end{nedalph}
\section{Small--$a$--expansion of the fermion self energy}\label{sec3}
\setcounter{equation}{0}
\hspace{1cm}
The action of the lattice $U\left(1\right)$ gauge theory we shall  consider is taken to be of the form
\begin{eqnarray}\label{SA1}
S[\overline{\psi},\psi;A;\{\epsilon_{\mu}\}]=S_{G}[A]+\sum\limits_{x}\overline{\psi}\left(x\right)\bigg[m-\frac{1}{a}\sum\limits_{\mu=1}^{D}\epsilon_{\mu}\gamma_{\mu}\left(U_{\epsilon_{\mu}\mu}T_{\epsilon_{\mu}\mu}-1\right)\bigg]\psi\left(x\right),
\end{eqnarray}
where $\sum\limits_{x}=\sum\limits_{n}a^{4}$, and  we have replaced the derivatives $\partial_{\mu}$ in the fermionic contribution to the continuum action by one--sided lattice differences \cite{stam} according to:
\begin{eqnarray*}
\partial_{\mu}\psi\left(x\right)&\rightarrow& \epsilon_{\mu}\ \ \frac{1}{a}\bigg\{\psi\left(x+a\epsilon_{\mu}\hat{\mu}\right)-\psi\left(x\right)\bigg\}\equiv \frac{1}{a}\epsilon_{\mu}\left(T_{\epsilon_{\mu}\mu}-1\right)\psi\left(x\right).
\end{eqnarray*}
Here $\hat{\mu}$ is a unit vector pointing in the $\mu$--direction, and $\epsilon_{\mu}=1$  ($\epsilon_{\mu}=-1$) corresponds to choosing the right (left) lattice difference. For $\epsilon_{\mu}=-1$ the link--variable $U_{-\mu}$   is defined by $U_{-\mu}\left(x\right)=U_{\mu}^{\dagger}\left(x-a\hat{\mu}\right)$. The contribution $S_{G}$ is the standard plaquette action for the gauge field.  From (\ref{SA1}) one readily obtains the following expression for the fermion propagator, the photon propagator in the Feynman gauge, and vertices,  coupling one or more gauge potentials to a $\overline{\psi}$--$\psi$ pair:

\pagebreak

\noindent{\it Fermion--Propagator:}

\begin{nedalph}\label{SA2a}
S_{F}\left(a,p,m,\{\epsilon_{\mu}\}\right)&=&\frac{m+\frac{2i}{a}\sum\limits_{\mu=1}^{4}\gamma_{\mu}\sin{\frac{p_{\mu}a}{2}}e^{i\epsilon_{\mu}p_{\mu}\frac{a}{2}}}
{m^{2}+\frac{4}{a^{2}}\sum\limits_{\rho=1}^{4}\sin^{2}{\frac{p_{\rho}a}{2}}e^{i\epsilon_{\rho}p_{\rho}a}}.
\end{eqnarray}
{\it Photon--Propagator:}
\begin{eqnarray}\label{SA2b}
D_{\mu\nu}\left(a,p\right)&=&\frac{\delta_{\mu\nu}}
{\frac{4}{a^{2}}\sum\limits_{\sigma}\sin^{2}\frac{p_{\sigma}a}{2}}.
\end{eqnarray}
{\it Vertices:}
\begin{eqnarray}\label{SA2c}
V_{\mu}\left(a,\overline{p},\{\epsilon_{\mu}\}\right)&=&-ig\gamma_{\mu}e^{i\epsilon_{\mu}\overline{p}_{\mu}a}, \nonumber\\
V_{\mu\mu_{1}}\left(a,\overline{p},\{\epsilon_{\mu}\}\right)&=&-a\epsilon_{\mu}g^{2}\delta_{\mu_{1}\mu}\gamma_{\mu}e^{i\epsilon_{\mu}\overline{p}_{\mu}a}, \nonumber\\
V_{\mu\mu_{1}\mu_{2}} \left(a,\overline{p},\{\epsilon_{\mu}\}\right) &=&+ia^{2}g^{3}\delta_{\mu_{2}\mu_{1}}\delta_{\mu_{1}\mu}\gamma_{\mu}e^{i\epsilon_{\mu}\overline{p}_{\mu}a}.\nonumber\\
\cdots&=&\cdots  
\end{nedalph}
Here  $\overline{p}_{\mu}=\frac{1}{2}\left(p^{in}+p^{out}\right)_{\mu}$, where $p_{\mu}^{in}$, ($p_{\mu}^{out}$) is the incoming (outgoing) momentum of the fermion.

\par
\begin{figure}[h]            
\leavevmode                     
\centering                      
\epsffile{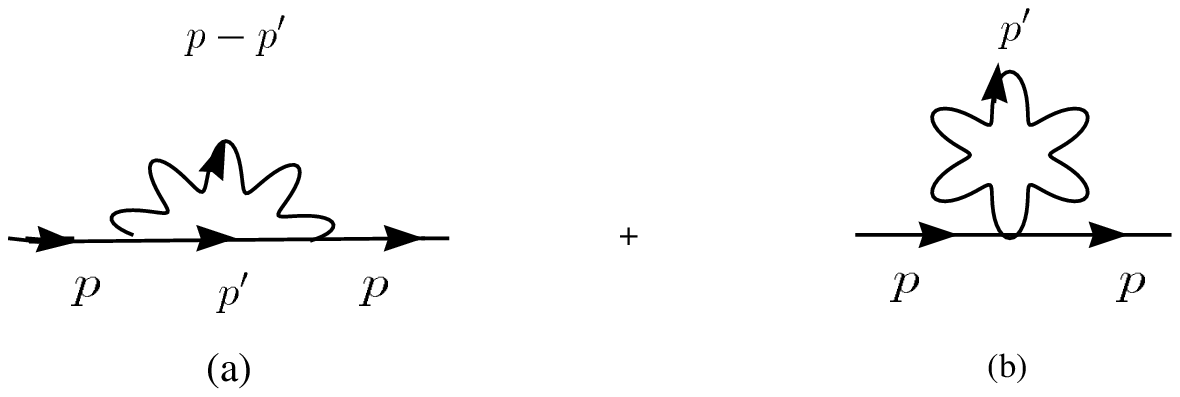}         
\caption{{\sl Diagrams contributing to the fermion self energy.}}\label{FIGONE}

\end{figure}

The fermion self energy receives a contribution from the two diagrams depicted in fig. \ref{FIGONE}.
The corresponding dimensionally regulated Feynman integrals are given by
\begin{nedalph}\label{SA3a}
\Sigma_{\left(\sigma\right)}\left(p,m,a,\{\epsilon_{\mu}\},D\right)&=& g^{2}\mu^{4-D}
\int\limits_{-\frac{\pi}{a}}^{+\frac{\pi}{a}}\ \frac{d^{D}p'}{\left(2\pi\right)^{D}}\  H^{\left(\sigma\right)}\left(p',p,m,a,\{\epsilon_{\mu}\} ,D\right),
\end{eqnarray}
where $\sigma=a, b$ and 
\begin{eqnarray}\label{SA3b}
\lefteqn{H^{\left(a\right)}\left(p',p,m,a,\{\epsilon_{\mu}\},D\right)=}\nonumber\\
&&-\ \frac{\sum\limits_{\nu=1}^{D}\bigg[\gamma_{\nu}e^{i\epsilon_{\nu}\frac{\left(p'+p\right)_{\nu}a}{2}}\left(m+\frac{2i}{a}\sum\limits_{\mu=1}^{D}\gamma_{\mu}\sin{\frac{p'_{\mu}a}{2}}e^{i\epsilon_{\mu}p'_{\mu}\frac{a}{2}}\right)\gamma_{\nu}e^{i\epsilon_{\nu}\frac{\left(p'+p\right)_{\nu}a}{2}}\bigg]}
{\bigg[\frac{4}{a^{2}} \sum\limits_{\tau=1}^{D} \sin^{2}{\frac{\left(p-p'\right)_{\tau}a}{2}}\bigg]\bigg
[m^{2}+\frac{4}{a^{2}}\sum\limits_{\rho=1}^{D}  \sin^{2}{\frac{p'_{\rho}a}{2}}e^{i\epsilon_{\rho}p'_{\rho}a}\bigg]},
\end{eqnarray}
and 
\begin{eqnarray}\label{SA3c}
H^{\left(b\right)}\left(p',p,m,a,\{\epsilon_{\mu}\},D\right)&=&-a\sum\limits_{\mu=1}^{D}\epsilon_{\mu}\gamma_{\mu}e^{i\epsilon_{\mu}p_{\mu}a}\frac{1}{\bigg[\frac{4}{a^{2}} \sum\limits_{\rho=1}^{D}\sin^{2}\frac{p'_{\rho}a}{2}\bigg]}. 
\end{nedalph}
In (\ref{SA3a}), $\mu$ is the usual arbitrary mass scale introduced  in the dimensional
 regularization scheme.
 
According to (\ref{A11a}) we have:                                                                                        
\begin{nedalph}\label{SA4a}
\lefteqn{\Sigma_{\left(\sigma\right)}\left(p,m,a,\{\epsilon_{\mu}\},D=4\right)=}\nonumber\\
&&=\lim\limits_{D\to 4}\bigg[
\Sigma_{\left(\sigma\right)}^{\left(\infty\right)} \left(p,m,a,\{\epsilon_{\mu}\},D\right ) +\left(a\mu\right)^{4-D}   \tilde{\Sigma}_{\left(\sigma\right)}\left(p,m,a,\{\epsilon_{\mu}\},D\right)
\bigg],
\end{eqnarray}
where  
 $ \Sigma_{\left(\sigma\right)}^{\left(\infty\right)}$ and  $ \tilde{\Sigma}_{\left(\sigma\right)}$   are defined by
\begin{eqnarray}\label{SA4b}
\lefteqn{\Sigma_{\left(\sigma\right)}^{\left(\infty\right)} \left(p,m,a,\{\epsilon_{\mu}\}, D\right )=}\nonumber\\
&& =g^{2}\mu^{4-D}
\int\limits_{-\infty}^{+\infty} \   \frac{d^{D}p'}{\left(2\pi\right)^{D}}\  T_{J} H^{\left(\sigma\right)}\left(p',p,m,a,\{\epsilon_{\mu}\},D\right) +{\cal{O}}\left(a^{J+1}\right),
\end{eqnarray}
and 
\begin{eqnarray}\label{SA4c}
\lefteqn{{\tilde{\Sigma}}_{\left(\sigma\right)} \left(p,m,a,\{\epsilon_{\mu}\},D\right )=}\nonumber\\
 &&=\frac{1}{a}\ g^{2}\ 
\int\limits_{-\pi}^{+\pi} \   \frac{d^{D}\hat{p'}}{\left(2\pi\right)^{D}}\  T_{J+1} H^{\left(\sigma\right)}\left(\hat{p'},ap,am,a=1,\{\epsilon_{\mu}\},D\right)+{\cal{O}}\left(a^{J+1}\right).
\end{nedalph}
Here $T_{l}H^{\left(\sigma\right)}$ denotes the Taylor expansion of $H^{\sigma}$ in the lattice spacing $a$ around $a=0$ up to order ${\cal{O}}\left(a^{l}\right)$. All integrals are to be calculated using dimensional regularization. In the limit $D\to 4$ the coefficients of $a^{j}$ in the expansion of (\ref{SA4b}) and  (\ref{SA4c}) will, in general, be ultraviolet (UV) and infrared (IR) divergent, respectively. The UV divergencies can be isolated in the standard way, since the integrals are of the continuum type. The technique for isolating the IR divergencies is described in the appendix. For $a\neq 0$, the UV and IR divergencies must cancel, since the original lattice Feynman integrals are finite.
\par
Consider first the leading contribution for $a\to 0$ to (\ref{SA4a}), which  is determined by the coefficient of ${\cal{O}}\left(a^{-1}\right)$ of (\ref{SA4c}).  This coefficient, which we denote by {\small$\tilde{\Sigma}^{\left(-1\right)}\left(  \{\epsilon_{\mu}\},D\right)$}, receives contributions from both diagrams in fig. \ref{FIGONE}. After making the change of integration variables $\epsilon_{\mu}\hat{p'}_{\mu}\to \hat{p'}_{\mu}$, one is led to the following non--covariant expression
\begin{nedalph}\label{SA5a}
\tilde{\Sigma}^{\left(-1\right)}\left(  \{\epsilon_{\mu}\},D\right) =\sum\limits_{\mu=1}^{D}\epsilon_{\mu}\gamma_{\mu}\ a_{\mu}\left(D\right), 
\end{eqnarray}
where
\begin{eqnarray}\label{SA5b}
a_{\mu}\left(D\right)&=&
-g^{2}\int\limits_{-\pi}^{+\pi}\frac{d^{D}\hat{p'}}{\left(2\pi\right)^{D}}\frac{1}
{ 4\sum\limits_{\rho=1}^{D}\sin^{2}\frac{\hat{p'}_{\rho}}{2}}
\bigg[
\frac{ 2i\sin\frac{\hat{p'}_{\mu}}{2} e^{\frac{i\hat{p'}_{\mu}}{2}}\left(2e^{i\hat{p'}_{\mu}}-\sum\limits_{\nu=1}^{D}e^{i\hat{p'}_{\nu}}\right)
}{4\sum\limits_{\rho=1}^{D}\sin^{2}\frac{\hat{p'}_{\rho}}{2}e^{i\hat{p'}_{\rho}}}+1
\bigg],\nonumber\\ 
\end{nedalph}
is IR--convergent.  Note that by averaging (\ref{SA5a}) over all possible $2^{D}$ sets of $\{\epsilon_{\mu}\}$  this expression is seen to vanish, as has already been pointed in ref. \cite{stam}.
\par
Next consider the contribution of ${\cal{O}}\left(a^{0}\right)$. In this order $\Sigma_{a}^{\left(\infty\right)}$ reduces to the usual continuum form of the Feynman integral for diagram (a) in fig. \ref{FIGONE}, while $\Sigma_{b}^{\left(\infty\right)}$ does not contribute in this order, as follows from (\ref{SA3c}). 
Hence $\Sigma^{\left(\infty\right)/\left(0\right)}$ is given by the well known dimensionally regulated continuum expression for the fermion self energy,
\begin{nedalph}\label{SA6a}
\Sigma^{\left(\infty\right)/\left(0\right)}\left(p\right)&=& {\cal{A}}^{\left(\infty\right)}\left(p\right) +
 \left(m-i\PS\right) {\cal{B}}^{\left(\infty\right)}\left(p\right),
\end{eqnarray}
where 
\begin{eqnarray}\label{SA6b}
{\cal{A}}^{\left(\infty\right)} \left(p\right)&=&\frac{-3mg^{2}}{8\pi^{2}}\ \frac{1}{\left(4-D\right)}+
\frac{3mg^{2}}{16\pi^{2}}\ln\frac{m^{2}}{\mu^{2}} \nonumber\\
&&+
\frac{mg^{2}}{8\pi^{2}}\int\limits_{0}^{1}
d\alpha \left(1+\alpha\right)\ \ln\bigg[\frac{ \alpha m^{2}+\alpha\left(1-\alpha\right)p^{2} }{\alpha^{2} m^{2}}\bigg]
+{\cal{C}}^{\left(\infty\right)}_{A},\\
{\cal{B}}^{\left(\infty\right)} \left(p\right)&=&\frac{-g^{2}}{8\pi^{2}}\ \frac{1}{\left(4-D\right)}+
\frac{g^{2}}{16\pi^{2}}\ln\frac{m^{2}}{\mu^{2}}  \nonumber\\
&&+\frac{g^{2}}{8\pi^{2}}\int\limits_{0}^{1}
d\alpha \left(1-\alpha\right)\ \ln\bigg[\frac{ \alpha m^{2}+\alpha\left(1-\alpha\right)p^{2} }{\alpha^{2} m^{2} }\bigg] +{\cal{C}}^{\left(\infty\right)}_{B}.\label{SA6c}
\end{eqnarray}
The constants ${\cal{C}}_{A}^{\left(\infty\right)}   $ and ${\cal{C}}_{B}^{\left(\infty\right)}   $ are given by
\begin{eqnarray}\label{SA6d}
{\cal{C}}_{A}^{\left(\infty\right)}   =
\frac{3mg^{2}}{16\pi^{2}}\gamma_{E}- \frac{5mg^{2}}{16\pi^{2}};&&
{\cal{C}}_{B}^{\left(\infty\right)} =
\frac{g^{2}}{16\pi ^{2}}\gamma_{E} -\frac{3g^{2}}{16\pi^{2}},
\end{nedalph}  
with $\gamma_{E}$ the Euler constant. 
\par
Consider next the contribution of ${\cal{O}}\left(a^{0}\right)$ to $\tilde{\Sigma}=
\tilde{\Sigma}_{a}+\tilde{\Sigma}_{b}$, where $\tilde{\Sigma}_{\left(\sigma\right)}$ ($\sigma=a, b$) has been defined in (\ref{SA4c}) with $H^{\left(\sigma\right)}$ given in (\ref{SA3b}, c). Both diagrams (a) and (b) contribute in this order. We denote this contribution by $  \tilde{\Sigma}^{\left(0\right)}  $. It is given by
\begin{nedalph}\label{SA7a}
 \tilde{\Sigma}^{\left(0\right)}  \left(p,m,\{\epsilon_{\mu}\},D\right) &=&
 \tilde{\Sigma}^{\left(0\right)}_{1} \left( p,m,\{\epsilon_{\mu}\},D \right) +
 \tilde{\Sigma}^{\left(0\right)}_{2}  \left(p,m,\{\epsilon_{\mu}\},D\right),
\end{eqnarray}
where
\begin{eqnarray}\label{SA7b}
{\tilde{\Sigma}}_{1}^{\left(0\right)} & =&\ 2g^{2}
\int\limits_{-\pi}^{+\pi}
\frac{d^{D}\hat{p'}}{\left(2\pi\right)^{D}}\ 
\frac{\sum\limits_{\mu=1}^{D}\epsilon_{\mu}\gamma_{\mu}\sin\frac{\hat{p'}_{\mu}}{2}e^{\frac{i\hat{p'}_{\mu}}{2}}\left(
2\epsilon_{\mu}p_{\mu}e^{i\hat{p'}_{\mu}} -\sum\limits_{\nu=1}^{D}
\epsilon_{\nu}p_{\nu}e^{i\hat{p'}_{\nu}}\right)
}
{
\bigg[4\sum\limits_{\rho=1}^{D}\sin^{2}\frac{ \hat{p'} _{\rho}}{2}\bigg] 
 \bigg[4\sum\limits_{\rho=1}^{D}\sin^{2}\frac{ \hat{p'} _{\rho}}{2}\ e^{i \hat{p'} _{\rho}}\bigg]   
}\nonumber\\
&&
-
ig^{2}\PS \int\limits_{-\pi}^{+\pi}
\frac{d^{D}\hat{p'}}{\left(2\pi\right)^{D}}\ 
\frac{1}{ \bigg[4\sum\limits_{\rho=1}^{D}\sin^{2}\frac{ \hat{p'} _{\rho}}{2}\bigg] } 
\end{eqnarray}
is infrared convergent for $D=4$, and 
\begin{eqnarray}\label{SA7c}
\tilde{\Sigma}_{2}^{\left(0\right)} &=&-mg^{2} \int\limits_{-\pi}^{+\pi}
\frac{d^{D}\hat{p'}}{\left(2\pi\right)^{D}}
\frac{\sum\limits_{\nu=1}^{D}e^{i\hat{p'}_{\nu}}}
{
\bigg[4\sum\limits_{\rho=1}^{D}\sin^{2}\frac{ \hat{p'} _{\rho}}{2}\bigg] 
 \bigg[4\sum\limits_{\rho=1}^{D}\sin^{2}\frac{ \hat{p'} _{\rho}}{2}\ e^{i \hat{p'} _{\rho}}\bigg]   
}
\nonumber\\
&&-4ig^{2}
\int\limits_{-\pi}^{+\pi}
\frac{d^{D}\hat{p'}}{\left(2\pi\right)^{D}}\ 
\frac{
\sum\limits_{\lambda=1}^{D}\epsilon_{\lambda}p_{\lambda}\sin\hat{p'}_{\lambda}\sum\limits_{\mu=1}^{D}\epsilon_{\mu}\gamma_{\mu}\sin\frac{\hat{p'}_{\mu}}{2}e^{\frac{i\hat{p'}_{\mu}}{2}}\left(
2e^{i\hat{p'}_{\mu}} -\sum\limits_{\nu=1}^{D}
e^{i\hat{p'}_{\nu}}\right)
}
{\bigg[4\sum\limits_{\rho=1}^{D}\sin^{2}\frac{ \hat{p'} _{\rho}}{2}\bigg] ^{2} \bigg[4\sum\limits_{\rho=1}^{D}\sin^{2}\frac{ \hat{p'} _{\rho}}{2}\ e^{i \hat{p'} _{\rho}}\bigg]   },\nonumber\\
\end{nedalph}
is infrared divergent for $D\to 4$. The  second term on the rhs of (\ref{SA7b}) is  the contribution from the diagram (b) in fig. \ref{FIGONE}. After some manipulations, expressions  (\ref{SA7b}) and (\ref{SA7c}) can be written in the form:
\begin{eqnarray}\label{SA8}
{\tilde{\Sigma}}_{i}^{\left(0\right)}\left( p,m,\{\epsilon_{\mu}\},D \right)&=& \tilde{{\cal{A}}}_{i}  +\left(m-i\PS\right)\tilde{{\cal{B}}}_{i}+ {\tilde{{\cal{C}}}}_{i}\left(p,\{\epsilon_{\mu}\},D\right),\ \ \ i=1,2.
\end{eqnarray}
Here $\tilde{{\cal{A}}}_{1} $ and $ \tilde{{\cal{B}}}_{1} $ are $\epsilon_{\mu}$--independent finite constants, involving lattice integrals which can only be computed numerically. The $\epsilon_{\mu}$--independent coefficients $\tilde{{\cal{A}}}_{2}  $ and $\tilde{{\cal{B}}}_{2}  $ are infrared divergent for $D\to 4$ and have the form
\begin{eqnarray*}
\tilde{{\cal{A}}}_{2} &=& -3mg^{2}\  {\cal{M}}\left(D\right) + {\tilde{\cal{D}}}_{A},\\
\tilde{{\cal{B}}}_{2} &=& -g^{2}\  {\cal{M}}\left(D\right) +{\tilde{\cal{D}}}_{B},
\end{eqnarray*}
where $ {\tilde{\cal{D}}}_{A}$ and $ {\tilde{\cal{D}}}_{B}$  are finite constants, and  ${\cal{M}}\left(D\right)$  is given by (see eq. (\ref{B5}) in the appendix),
\begin{eqnarray}\label{SA9}
{\cal{M}}\left(D\right) \equiv\int\limits_{-\pi}^{+\pi}
\frac{d^{D}\hat{p'}}{\left(2\pi\right)^{D}}
\frac{1}{
\bigg[4\sum\limits_{\rho=1}^{D}\sin^{2}\frac{ \hat{p'} _{\rho}}{2}\bigg] 
^{2}}  =\frac{-1}{8\pi^{2}}\ \frac{1}{\left(4-D\right)}+ {\tilde{\cal{N}}}_{2},
\end{eqnarray}
with  $ {\tilde{\cal{N}}}_{2}$ a numerical constant. 
\\
The $\epsilon_{\mu}$--dependent part in (\ref{SA8}) has the following  non--covariant structure:
\begin{nedalph} 
 \lefteqn{ \tilde{{\cal{C}}}_{1}\left(p,\{\epsilon_{\mu}\}, D\right) =}\nonumber\\
&=&-2g^{2}\ \sum\limits_{\stackrel{\mu,\nu=1}{\mu\neq\nu}}^{D}
\epsilon_{\mu}\epsilon_{\nu}
\gamma_{\mu}p_{\nu}
\int\limits_{-\pi}^{+\pi}
\frac{d^{D}\hat{p'}}{\left(2\pi\right)^{D}}
\frac{\sin\frac{\hat{p'}_{\mu}}{2}\ e^{\frac{i\hat{p'}_{\mu}}{2}}\ e^{i\hat{p'}_{\nu}}
}
{
\bigg[4\sum\limits_{\rho=1}^{D}\sin^{2}\frac{ \hat{p'} _{\rho}}{2}\bigg] 
 \bigg[4\sum\limits_{\rho=1}^{D}\sin^{2}\frac{ \hat{p'} _{\rho}}{2}\ e^{i \hat{p'} _{\rho}}\bigg]   },\label{SA10a}\\
\lefteqn{
\tilde{{\cal{C}}}_{2}\left(p,\{\epsilon_{\mu}\}, D\right) 
=}\nonumber\\
&=&-4ig^{2}
 \sum\limits_{\stackrel {\mu,\nu=1} {\mu\neq \nu}}^{D} \epsilon_{\mu}\epsilon_{\nu}\gamma_{\mu}p_{\nu}
\int\limits_{-\pi}^{+\pi}
\frac{d^{D}\hat{p'}}{\left(2\pi\right)^{D}}\frac{ 
\sin\hat{p'}_{\nu}\sin\frac{\hat{p'}_{\mu}}{2}e^{\frac{i\hat{p'}_{\mu}}{2}}\left(
2e^{i\hat{p'}_{\mu}}-\sum\limits_{\lambda=1}^{D}e^{i\hat{p'}_{\lambda}}
\right)
}{\bigg[4\sum\limits_{\rho=1}^{D}\sin^{2}\frac{ \hat{p'} _{\rho}}{2}\bigg] ^{2}
 \bigg[4\sum\limits_{\rho=1}^{D}\sin^{2}\frac{ \hat{p'} _{\rho}}{2}\ e^{i \hat{p'} _{\rho}}\bigg]   
}. \label{SA10b}
\end{nedalph}While (\ref{SA10a}) is finite for $D=4$, (\ref{SA10b}) is infrared divergent for $D\to 4$.
\par\noindent
From  (\ref{SA7a})--(\ref{SA10b}) we therefore find that, in this order  the contribution of the second term on the rhs of  (\ref{SA4a}) to $\Sigma=\Sigma_{\left(a\right)}+\Sigma_{\left(b\right)}$ is given by:
\begin{nedalph}\label{SA11a}
\left(a\mu\right)^{4 -D}\tilde{\Sigma}^{\left(0\right)}\left(p,m,\{\epsilon_{\mu}\},D\right)
&\stackrel{D\to4}{\approx}&
\tilde{{\cal{A}}} +\left(m-i\PS\right)  \tilde{{\cal{B}}} +\tilde{{\cal{C}}}\left(p,\{\epsilon_{\mu}\}, D\right) ,
\end{eqnarray}
where
\begin{eqnarray}
  \tilde{{\cal{A}}}& =& 
\frac{3mg^{2}}{8\pi^{2}}\ \frac{1}{\left(4-D\right)} +
\frac{3mg^{2}}{8\pi^{2}}\ \ln\left(a\mu\right)+  \tilde{{\cal{C}}}_{A},
\label{SA11b}\\
 \tilde{{\cal{B}}} &=&
\frac{g^{2}}{8\pi^{2}}\ \frac{1}{\left(4-D\right)} +
\frac{g^{2}}{8\pi^{2}}\ \ln\left(a\mu\right)+\tilde{{\cal{C}}}_{B}. \label{SA11c}
\end{nedalph}
Here $ \tilde{{\cal{C}}}_{A}$ and $ \tilde{{\cal{C}}}_{B}$ are finite constants, and $\tilde{{\cal{C}}}\left(p,\{\epsilon_{\mu}\}, D\right)=\sum\limits_{i=1}^{2}
\tilde{{\cal{C}}}_{i}\left(p,\{\epsilon_{\mu}\}, D\right)$ (see eqs. (\ref{SA10a}, b)). Note that the expressions (\ref{SA11b}, c) involve two regulators: the lattice spacing $a$ and the dimension $D$. 
\par\noindent
Combining the contributions  (\ref{SA5a}), (\ref{SA6a}) and (\ref{SA11a})
 we arrive the following expression for the fermion self energy for $D=4$,
\begin{nedalph}\label {SA12a}
\lefteqn{\Sigma\left(p,m,a,\{\epsilon_{\mu}\},D=4\right)=}\nonumber\\
&&=\frac{1}{a}\ \sum\limits_{\mu=1}^{4}\epsilon_{\mu}\gamma_{\mu}a_{\mu}+\sum\limits_{\stackrel{\mu,\nu=1}{\mu\neq\nu}}^{4}\epsilon_{\mu}\epsilon_{\nu}\gamma_{\mu}p_{\nu}b_{\mu\nu}+{\cal{A}}+\left(m-i\PS\right){\cal{B}} +{\cal{O}}\left(a\right),
\end{eqnarray}
where $a_{\mu}\equiv a_{\mu}\left(D=4\right)$ and the (for $D\to 4$ infrared divergent) coefficients $b_{\mu\nu}$ are obtained by combining (\ref{SA10a}) and (\ref{SA10b}). The expressions for  ${\cal{A}}$ and ${\cal{B}}$ are given by, 
\newpage
\begin{eqnarray}\label{SA12b}
{\cal{A}}&=&\delta m+\frac{mg^{2}}{8\pi^{2}}\int\limits_{0}^{1}
d\alpha \left(1+\alpha\right)\ \ln\bigg[\frac{ \alpha m^{2}+\alpha\left(1-\alpha\right)p^{2} }{\alpha^{2}m^{2} }\bigg], \nonumber\\
{\cal{B}}&=&\left(1-{\cal{Z}}_{2}^{-1}\right)+\frac{g^{2}}{8\pi^{2}}\int\limits_{0}^{1}
d\alpha \left(1-\alpha\right)\ \ln\bigg[\frac{ \alpha m^{2}+\alpha\left(1-\alpha\right)p^{2} }{\alpha^{2} m^{2} }\bigg],
\end{eqnarray}
 where
\begin{eqnarray}\label{SA12c}
\delta m&=&\frac{3mg^{2}}{8\pi^{2}}\ln \left(am\right)+ {\cal{C}}_{m}, \nonumber\\
{\cal{Z}}_{2}^{-1}&=&1- \frac{g^{2}}{8\pi^{2}}\ln \left(am\right)+{\cal{C}}_{z}.
\end{nedalph}
Here $ {\cal{C}}_{m} ={\cal{C}}_{A}^{\left(\infty\right)} +\tilde{\cal{C}}_{A}  $ and $ {\cal{C}}_{z} ={\cal{C}}_{B}^{\left(\infty\right)} +\tilde{\cal{C}}_{B}  $, are the $\epsilon_{\mu}$--independent constants. These constants differ from those obtained in the dimensional regularization scheme for continuum Feynman integrals, since we  have used the lattice as a regulator.
\par
Note that a remarkable cancellation of $\epsilon_{\mu}$--independent, but in the limit $D$ to $4$, infrared and ultraviolet divergent terms has occured after combining the above mentioned contribution, leaving us with a $\mu$--independent expression with the lattice spacing as the only regulator!
\par
The first two terms in (\ref {SA12a}) cannot be eliminated by introducing a renormalized mass parameter, and renormalized fields. Both terms have a non--covariant structure for any choice of $\{\epsilon_{\mu}\}$. These non--covariant contributions are  seen to vanish if they are averaged over all possible sets of $\{\epsilon_{\mu}\}$. Such an averaging procedure had been proposed in ref. \cite{stam}, where the fermion self energy has been discussed on a qualitative level. 
\section{Small--$a$--expansion of the Vacuum polarization tensor and vertex function}\label{sec4}
\setcounter{equation}{0}
\hspace{1cm}We begin with the discussion of the small--$a$--expansion of the one loop contribution to the vacuum polarization tensor.

\newpage

\noindent{\it i ) Vacuum polarization tensor}.

\bigskip

\par
\begin{figure}[h]            
\leavevmode                     
\centering                      
\epsffile{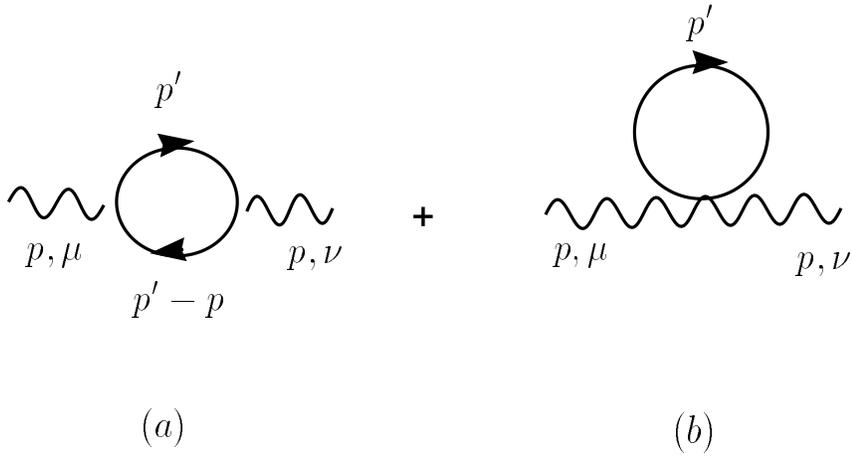}         
\caption  
        {{{\sl Diagrams contributing to the vacuum polarization tensor.}}
}\label{FIGTWO}      

\end{figure}

The diagrams contributing to the vacuum polarization are shown in fig. \ref{FIGTWO}, of which diagram (b) has no continuum analog. 
The corresponding Feynman integrals have the form: 
\begin{eqnarray}\label{SB1}
\Pi^{\mu\nu}_{\left(\sigma\right)}\left(p,m,a,\{\epsilon_{\mu}\},D\right)&=& g^{2}\mu^{4-D}
 \int\limits_{-\frac{\pi}{a}}^{+\frac{\pi}{a}}\ \frac{d^{D}p'}{\left(2\pi\right)^{D}}\  H^{\mu\nu/\left(\sigma\right)}\left(p',p,m,a,\{\epsilon_{\mu}\} ,D\right),\ \ \sigma=a,b, \nonumber\\
\end{eqnarray}
where the integrand $H^{\mu\nu/\left(\sigma\right)}$ can be readily obtained using the expressions for the propagator and vertices given in (\ref{SA2a}--c).        
The small--$a$--expansion of $\Pi_{\left(\sigma\right)}^{\mu\nu}$
analogous to (\ref{SA4a}--c) now reads:
\begin{nedalph}\label{SB2a}
\lefteqn{
\Pi^{\mu\nu}_{\left(\sigma\right)}\left(p,m,a,\{\epsilon_{\mu}\},D=4\right)=
}\nonumber\\
&&
= \lim\limits_{D\to 4}\bigg[
\Pi_{\left(\sigma\right)}^{\mu\nu/\left(\infty\right)} \left(p,m,a,\{\epsilon_{\mu}\},D\right ) +\left(a\mu\right)^{4-D}   \tilde{\Pi}^{\mu\nu}_{\left(\sigma\right)}\left(p,m,a,\{\epsilon_{\mu}\},D\right)
\bigg],
\end{eqnarray}
where  
 $ \Pi_{\left(\sigma\right)}^{\mu\nu/\left(\infty\right)}$ and $ \tilde{\Pi}^{\mu\nu}_{\left(\sigma\right)}$   are defined by
\begin{eqnarray}\label{SB2b}
\lefteqn{
\Pi_{\left(\sigma\right)}^{\mu\nu/\left(\infty\right)} \left(p,m,a,\{\epsilon_{\mu}\}, D\right ) =
}\nonumber\\
&&
=g^{2}\mu^{4-D}
\int\limits_{-\infty}^{+\infty} \   \frac{d^{D}p'}{\left(2\pi\right)^{D}}\  T_{J} H^{\mu\nu/\left(\sigma\right)}\left(p',p,m,a,\{\epsilon_{\mu}\},D\right) +{\cal{O}}\left(a^{J+1}\right),
\end{eqnarray}
and 
\begin{eqnarray}\label{SB2c}
\lefteqn{
\tilde{\Pi}^{\mu\nu}_{\left(\sigma\right)} \left(p,m,a,\{\epsilon_{\mu}\},D\right ) =
}\nonumber\\
&&
=
\frac{1}{a^{2}}g^{2} 
\int\limits_{-\pi}^{+\pi}\frac{d^{D}\hat{p'}}{\left(2\pi\right)^{D}}\ T_{J+2} H^{\mu\nu/\left(\sigma\right)}\left(\hat{p'},ap,am,a=1,\{\epsilon_{\mu}\},D\right) +{\cal{O}}\left(a^{J+1}\right).
\end{nedalph}
The leading contributions for $a\to 0$  to (\ref{SB2a}) is determined by the coefficient of ${\cal{O}}\left(a^{-2}\right)$ and ${\cal{O}}\left(a^{-1}\right)$ of $\tilde{\Pi}^{\mu\nu}_{\left(\sigma\right)}$, which we denote by $\tilde{\Pi}^{\mu\nu/\left(-2\right)}_{\left(\sigma\right)}$ and 
$\tilde{\Pi}^{\mu\nu/\left(-1\right)}_{\left(\sigma\right)}$. The coefficient of 
$\tilde{\Pi}^{\mu\nu/\left(-2\right)}_{\left(\sigma\right)}$ for $\sigma=a$ and $\sigma=b$  is given by   
\begin{nedalph}\label{SB3a}
 \tilde{\Pi}_{a}^{\mu\nu/\left(-2\right)}\left(\{\epsilon_{\mu}\},D\right)
&=&-32g^{2}\epsilon_{\mu}\epsilon_{\nu}\int\limits_{-\pi}^{+\pi}\frac{d^{D}\hat{p'}}
{\left(2\pi\right)^{D}}  
 \ \frac{
\sin\frac{\hat{p'}_{\mu}}{2}\sin\frac{\hat{p'}_{\nu}}{2}
e^{\frac{3}{2}i\hat{p'}_{\mu}}e^{\frac{3}{2}i\hat{p'}_{\nu}}
}{\bigg[4\sum\limits_{\rho=1}^{D}\sin^{2}\frac{\hat{p'}_{\rho}}{2}e^{i\hat{p'}_{\rho}}\bigg]^{2}}\nonumber\\
&&+
4g^{2}   \delta_{\mu\nu}\int\limits_{-\pi}^{+\pi}\frac{d^{D}\hat{p'}}{\left(2\pi\right)^{D}} \ \frac{ e^{2i\hat{p'}_{\mu}}}{\bigg[4\sum\limits_{\rho=1}^{D}\sin^{2}\frac{\hat{p'}_{\rho}}{2}e^{i\hat{p'}_{\rho}}\bigg] },
\end{eqnarray}
and 
\begin{eqnarray}\label{SB3b}
\tilde{\Pi}_{b}^{\mu\nu/\left(-2\right)}\left(D\right)&=& 8ig^{2}\delta_{\mu\nu}\ \int\limits_{-\pi}^{+\pi}\ \frac{d^{D}\hat{p'}}{\left(2\pi\right)^{D}}\ \frac{\sin\frac{\hat{p'}_{\mu}}{2}e^{\frac{3}{2}i\hat{p'}_{\mu}}}
{
\bigg[4\sum\limits_{\rho=1}^{D}\sin^{2}\frac{\hat{p'}_{\rho}}{2}e^{i\hat{p'}_{\rho}}\bigg]
}.
\end{nedalph}
After a partial integration of the first term in (\ref{SB3a}) one finds that
\begin{eqnarray}\label{SB4}
\tilde{\Pi}_{a}^{\mu\nu/\left(-2\right)} = - \tilde{\Pi}_{b}^{\mu\nu/\left(-2\right)}.
\end{eqnarray}
Hence there is no quadratic divergence (for $a\to 0$), as expected from gauge invariance. The lattice expression for the coefficients $\tilde{\Pi}^{\mu\nu/\left(-1\right)}_{\left(\sigma\right)}$ is rather lengthy and we do not present it here. After some trigonometric manipulations and partial integrations we find that 
\begin{eqnarray}\label{SB5}
\tilde{\Pi}^{\mu\nu/\left(-1\right)}_{\left(a\right)}=\tilde{\Pi}^{\mu\nu/\left(-1\right)}_{\left(b\right)}=0.
\end{eqnarray} 
Hence there is also no linearly divergent contribution to $\Pi^{\mu\nu}$ for $a\to 0$. Note  that (\ref{SB4}) and (\ref{SB5}) hold for arbitrary choice of $\{\epsilon_{\mu}\}$.
\par
Consider next the contribution of ${\cal{O}}\left(a^{0}\right)$ to $\Pi^{\mu\nu/\left(\infty\right)}_{\sigma}$ in eq. (\ref{SB2a}). For $\sigma=a$ it is given by the usual dimensionally regulated continuum expression for the vacuum polarization tensor, while $\Pi^{\mu\nu/\left(\infty\right)}_{b}$ does not contribute:
\begin{nedalph}\label{SB6a}
 [\Pi^{\mu\nu/\left(\infty\right)}]_{{\cal{O}}\left(a^{0}\right)}=
\left(p_{\mu}p_{\nu}-p^{2}\delta_{\mu\nu}\right)\Pi^{\left(\infty\right)/\left(0\right)}\left(p,m,D\right),
\end{eqnarray}
where
\begin{eqnarray}\label{SB6b}
\lefteqn{\Pi^{\left(\infty\right)/\left(0\right)}\left(p,m,D\right)
=\frac{g^{2}}{6\pi^{2}}\ \frac{1}{\left(4-D\right)}
-\frac{g^{2}}{12\pi^{2}}\ \ln\frac{m^{2}}{\mu^{2}}
}\nonumber\\
&&
-\frac{g^{2}}{2\pi^{2}}\ \int\limits_{0}^{1}\alpha\left(1-\alpha\right)\ \ln\bigg[\frac{m^{2}+\alpha\left(1-\alpha\right)p^{2}}{m^{2}}\bigg]\ d\alpha
+{\cal{C}}^{\left(\infty\right)},
\end{nedalph}
and 
$
{\cal{C}}^{\left(\infty\right)}=\frac{-g^{2}}{12\pi^{2}} \ \gamma_{E}
$. 
\par
The expression for the contribution of ${\cal{O}}\left(a^{0}\right)$ to $ \tilde{\Pi}^{\mu\nu} = \tilde{\Pi}^{\mu\nu}_{\left(a\right)}+\tilde{\Pi}^{\mu\nu}_{\left(b\right)} $ is very lengthy. After some work (involving mainly partial integrations) one finds
\begin{eqnarray}\label{SB7}
\tilde{\Pi}^{\mu\nu/\left(0\right)}\left(p,D\right)&=&-\frac{4}{3}g^{2} \delta_{\mu\nu}\sum\limits_{\lambda=1}^{D}p_{\lambda}^{2} \ \int\limits_{-\pi}^{+\pi}\frac{d^{D}\hat{p'}}{\left(2\pi\right)^{D}}\ \frac{e^{2i\hat{p'}_{\mu}}e^{2i\hat{p'}_{\lambda}}
}{
\bigg[4\sum\limits_{\rho=1}^{D}\sin^{2}{\frac{\hat{p'}_{\rho}}{2}}e^{i\hat{p'}_{\rho}}\bigg]   
^{2}}\nonumber\\
&&+\frac{4}{3}g^{2}p_{\mu}p_{\nu}\ \int\limits_{-\pi}^{+\pi}\frac{d^{D}\hat{p'}}{\left(2\pi\right)^{D}}\ \frac{e^{2i\hat{p'}_{\mu}}e^{2i\hat{p'}_{\nu}}
}{
\bigg[4\sum\limits_{\rho=1}^{D}\sin^{2}{\frac{\hat{p'}_{\rho}}{2}}e^{i\hat{p'}_{\rho}}\bigg]   
^{2}}
\end{eqnarray}
This contribution can be shown to be of the form 
\begin{nedalph}\label{SB8a}
\tilde{\Pi}^{\mu\nu/\left(0\right)}\left(p,D\right)=
\left(p_{\mu}p_{\nu}-p^{2}\delta_{\mu\nu}\right)\tilde{\Pi}^{\left(0\right)}\left(D\right).
\end{eqnarray}
To this effect one must consider separately the cases when the momenta appearing in the exponentials of the integrands carry the same or different indices. Making use of the fact that for $\mu\neq \lambda$ the integrals in (\ref{SB7}) are
independent of the choice of $\mu$ and $\lambda$, one finds that 
\begin{eqnarray}\label{SB8b}
\tilde{\Pi}^{\left(0\right)}\left(D\right) &=&\frac{4}{3}g^{2}  {\cal{M}}\left(D\right)
+\bigg[\tilde{\Pi}^{\left(0\right)} \left(D\right) \bigg]_{\mbox{\small{reg.}}}, 
\end{eqnarray}
where ${\cal{M}}\left(D\right)$ is given by (\ref{SA9}). The second term in (\ref{SB8b}) is regular in the limit $D\to 4$, and reads 
\begin{eqnarray}\label{SB8c}
\lefteqn{
\bigg[\tilde{\Pi}^{\left(0\right)} \left(D=4 \right) \bigg]_{\mbox{\small{reg.}}}
=}\nonumber\\
&&=\frac{4}{3}g^{2}\int\limits_{-\pi}^{+\pi}\frac{d^{4}\hat{p'}}{\left(2\pi\right)^{4}}\ \Biggl[\ \frac{1}{12}
\sum\limits_{\stackrel{\mu,\sigma=1}{\mu\neq\sigma}  } ^{4}\frac{
e^{2i\hat{p'}_{\mu}}e^{2i\hat{p'}_{\sigma}}
}{
\bigg[
4\sum\limits_{\rho=1}^{4}\sin^{2}\frac{\hat{p'}_{\rho}}{2}e^{i\hat{p'}_{\rho}}
\bigg]^{2}
}   -
\frac{1}{
\bigg[
4\sum\limits_{\rho=1}^{4}\sin^{2}\frac{\hat{p'}_{\rho}}{2}
\bigg]^{2}
} \Biggl].
\end{nedalph}
We therefore find that the contribution of the second term in (\ref{SB2a}), in this order, is given by 
\begin{eqnarray}\label {SB9}
\left(a\mu\right)^{4-D} \tilde{\Pi}^{\mu\nu/\left(0\right)}\left(p,D\right)\stackrel{D\to 4}{\approx}\left(p_{\mu}p_{\nu}-p^{2}\delta_{\mu\nu}\right)\bigg[\frac{-g^{2}}{6\pi^{2}}\ \frac{1}{\left(4 -D\right)}-\frac{g^{2}}{6\pi^{2}}\ln\left(a\mu\right)+\tilde{\cal{C}}\bigg],
\end{eqnarray}
where $\tilde{\cal{C}}$ is a finite constant. Notice again that this expression involves two regulators: the lattice spacing and the dimension $D$. By combining (\ref{SB9}) with (\ref{SB6a}) the $D$--dependent terms are seen to cancel, and we are left with a $\mu$--independent expression valid up to order ${\cal{O}}\left(a^{0}\right)$,  in which the lattice spacing appears as the only regularization parameter:
\begin{nedalph}\label{SB10a}
\Pi^{\mu\nu}=   \left(p_{\mu}p_{\nu}-p^{2}\delta_{\mu\nu}\right)  \Pi^{\left(0\right)}  \left(p\right) +{\cal{O}}\left(a\right).
\end{eqnarray}
Here
\begin{eqnarray}\label{SB10b}
\Pi^{\left(0\right)}  \left(p\right)=\left({\cal{Z}}_{3}^{-1}-1\right)-
 \frac{g^{2}}{2\pi^{2}}\ \int\limits_{0}^{1}\alpha\left(1-\alpha\right)\ \ln\bigg[\frac{m^{2}+\alpha\left(1-\alpha\right)p^{2}}{m^{2}}\bigg]\ d\alpha, 
\end{eqnarray} 
and 
\begin{eqnarray}\label{SB10c}
{\cal{Z}}_{3}^{-1}-1 &=&
 -\frac{g^{2}}{6\pi^{2}}\ln\left(am\right)+ {\cal{C}},
\end{nedalph}
where $ {\cal{C}}={\cal{C}}^{\left(\infty\right)}+\tilde{\cal{C}} $ is a finite $\epsilon_{\mu}$--independent constant. Notice that in contrast to the case of the fermion self energy (\ref {SA12a}), where the non--covariant contributions of ${\cal{O}}\left(a^{-1}\right)$ and ${\cal{O}}\left(a^{0}\right)$ could only be eliminated by averaging the expression over all possible sets of $\{\epsilon_{\mu}\}$, the expressions (\ref {SB10a}--c)  have the correct continuum structure for any choice of $\{\epsilon_{\mu}\}$. The expected cancellation of the $D$--dependent terms, observed above, has been checked up to ${\cal{O}}\left(a^{2}\right)$\footnote{{\normalsize{Only one of us (N.S.) had the nerve to carry out  this very extensive computation.}}}. 
\bigskip

\par
\noindent{\it {ii) Vertex function}}.

\bigskip
\par
\begin{figure}[h]            
\leavevmode                     
\centering                      
\epsffile{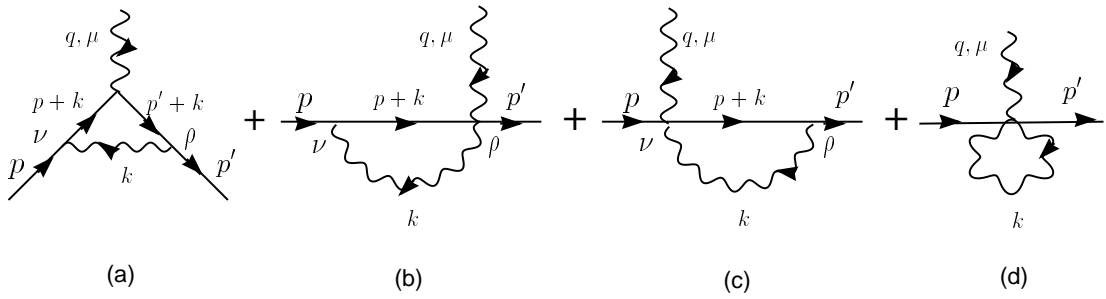}         
\caption       
  {{{\sl Diagrams contributing to the vertex function. }}
 }      
\label{FIGTHREE}
\end{figure}

In analogy to (\ref{SA2c}) we write the vertex function in the form:
\begin{eqnarray*}
V_{\mu}^{\left(1\right)}\left(p,p',m;a,\{\epsilon_{\mu}\},D\right)&=&
-ig\left(\gamma_{\mu}+\Lambda_{\mu}\left(p,p',m;a,\{\epsilon_{\mu}\},D\right)\right)\ e^{\frac{i\epsilon_{\mu}  \left(p+p'\right)_{\mu}}{2} },
\end{eqnarray*}
In one--loop order $\Lambda_{\mu}$ receives a contribution from the diagrams shown in fig. \ref{FIGTHREE}.
The lattice Feynman integrals contributing to $\Lambda_{\mu}$, which we denote by $\Lambda_{\mu}^{\left(\sigma\right)}$ ($\sigma=a,b,c,d$) can be readily written down using the expression (\ref{SA2a}--c) for the propagators and vertices. The small--$a$--expansion of $\Lambda_{\mu}=\sum\limits_{\sigma}\Lambda_{\mu}^{\left(\sigma\right)}$ now reads:
\begin{eqnarray}\label{SB11}
\lefteqn{
\Lambda_{\mu}\left(p,p',m;a,\{\epsilon_{\mu}\},D=4\right) =}\nonumber\\
&&=\lim\limits_{D\to 4}\bigg[
\Lambda_{\mu}^{\left(\infty\right)} \left(p,p',m; a,\{\epsilon_{\mu}\},D\right ) +\left(a\mu\right)^{4-D}   \tilde{\Lambda}_{\mu}\left(p,p',m;a,\{\epsilon_{\mu}\},D\right)
\bigg],
\end{eqnarray}
with $\Lambda_{\mu}^{\left(\infty\right)}=\sum\limits_{\sigma}\Lambda_{\mu}
^{\left(\sigma\right)/\left(\infty\right)}$ and $\tilde{\Lambda}_{\mu}=\sum\limits_{\sigma} \tilde{\Lambda}_{\mu}^{\left(\sigma\right)} $. Here $\Lambda_{\mu}^{\left(\sigma\right)/\left(\infty\right)}$ and $\tilde{\Lambda}_{\mu}^{\left(\sigma\right)} $ are defined by expressions analogous to (\ref{SA4b}) and (\ref{SA4c}), respectively, except that the factor $a^{-1}$ in (\ref{SA4c}) is replaced by $1$, and $T_{J+1}$  is replaced by $T_{J}$. The leading term in their expansions in $a$ is  of order ${\cal{O}}\left(a^{0}\right)$. In this order $ \Lambda_{\mu}^{\left(a\right)/\left(\infty\right)} $ is given by the usual $D$--dimensionally regulated continuum expression, while $ \Lambda_{\mu}^{\left(\sigma\right)/\left(\infty\right)}$ vanishes for $\sigma=b,c,d$ in this order. Hence
\begin{eqnarray}\label{SB12}
\Lambda_{\mu}^{\left(\infty\right)/\left(0\right)}&=& \frac{g^{2}}{8\pi^{2}}\ \gamma_{\mu}\ \frac{1}{\left(4-D\right)}-\frac{g^{2}}{16\pi^{2}}\ \gamma_{\mu}\ln\frac{m^{2}}{\mu^{2}}+ \Lambda_{\mu}^{\left(\mbox{\small{reg.}}\right)},
\end{eqnarray}
where $\Lambda_{\mu}^{\left(\mbox{\small{reg.}}\right)}$ is the known regular contribution of the continuum formulation. 
\par
Next consider the contribution of ${\cal{O}}\left(a^{0}\right)$ to $\tilde{\Lambda}_{\mu}$. All four diagrams in fig.  \ref{FIGTHREE} contribute in this case. The expression for $\tilde{\Lambda}_{\mu}^{\left(0\right)}$ is rather lengthy, and we will not give it here explicitely, since it can be obtained in a straightforward way.  After some simple algebra, one find that $\tilde{\Lambda}_{\mu}^{\left(0\right)}$ has the following momentum independent form:
\begin{eqnarray}\label{SB13}
\tilde{\Lambda}_{\mu}^{\left(0\right)}\left(\{\epsilon_{\mu}\},D\right)&=&
\frac{-g^{2}}{8\pi^{2}}\ \gamma_{\mu}\ \frac{1}{\left(4-D\right)}+\gamma_{\mu}\tilde{\cal{D}}+\tilde{\cal{C}}\left(\{\epsilon_{\mu}\},D\right).
\end{eqnarray}
The first term is entirely determined by diagram (a). The second term, which receives contributions from all four diagrams,  is finite for $D\to 4$. Finally, the third term, appearing on the rhs of (\ref{SB13}), is a non--covariant,  $\epsilon_{\mu}$--dependent term. It is given by
\begin{eqnarray}\label{SB14}
\lefteqn{\tilde{\cal{C}}_{\mu} \left(\{\epsilon_{\mu}\}, D\right)=}\nonumber\\
  &=&
+4g^{2}
\sum\limits_{\stackrel{\nu,\tau,\theta=1}{\tau\neq\theta}}^{D} \epsilon_{\theta}\ \epsilon_{\tau} \gamma_{\nu}\gamma_{\tau}\gamma_{\mu}\gamma_{\theta}\gamma_{\nu}
\int\limits_{-\pi}^{+\pi}\frac{d^{D}\hat{k}}{\left(2\pi\right)^{D}}\ \frac{
e^{i\hat{k}_{\mu}} 
e^{i\hat{k}_{\nu}}\left(\sin\frac{\hat{k}_{\tau}}{2}\sin\frac{\hat{k}_{\theta}}{2}e^{\frac{i\left(\hat{k}_{\tau}+\hat{k}_{\theta}\right)}{2}}\right)
}
{\bigg[4\sum\limits_{\rho=1}^{D}\sin^{2}\frac{\hat{k}_{\rho}}{2}\bigg]\bigg[4\sum\limits_{\rho=1}^{D}\sin^{2}\frac{\hat{k}_{\rho}}{2}e^{i\hat{k}_{\rho}}\bigg]^{2}}
\nonumber\\
&&
+4ig^{2}\sum\limits_{\stackrel {\tau=1} {\mu\neq\tau} }^{D}\epsilon_{\mu}\epsilon_{\tau}\gamma_{\tau}\ \int\limits_{-\pi}^{+\pi}\frac{d^{D}\hat{k}}{\left(2\pi\right)^{D}}\ 
\frac{
\sin\frac{\hat{k}_{\tau}}{2}e^{\frac{i\hat{k}_{\tau}}{2}}e^{i\hat{k}_{\mu}}
}
{\bigg[4\sum\limits_{\rho=1}^{D}\sin^{2}\frac{\hat{k}_{\rho}}{2}\bigg]\bigg[4\sum\limits_{\rho=1}^{D}\sin^{2}\frac{\hat{k}_{\rho}}{2}e^{i\hat{k}_{\rho}}\bigg]},
\end{eqnarray}
where the first term arises from diagram (a), and the second term from (b) and  (c). 
This expression can be reduced to the form
\begin{eqnarray}\label{SB15}
\tilde{\cal{C}}_{\mu}\left(\{\epsilon_{\mu}\},D\right)&=&
\sum\limits_{\stackrel{\nu,\tau,\theta=1}{\tau\neq\theta}}^{D}\epsilon_{\theta}\epsilon_{\tau}\gamma_{\mu}\gamma_{\tau}\gamma_{\theta}\ a_{\mu\nu\tau\theta}\left(D\right)+\sum\limits_{\stackrel {\tau=1} {\mu\neq\tau} }^{D}\epsilon_{\mu}\epsilon_{\tau}\gamma_{\tau}\ b_{\mu\tau}\left(D\right),
\end{eqnarray}
where the coefficients $a_{\mu\nu\tau\theta}\left(D\right)$ are given, in the limit $D\to 4$, by infrared divergent integrals, while $b_{\mu\tau}\left(D\right)$ are finite constants. From (\ref{SB13}) we see that the second term appearing within square brackets in 
(\ref{SB11}) takes the following form for $D\to 4$:
\begin{eqnarray}\label{SB16}
\left(a\mu\right)^{4-D}\tilde{\Lambda}_{\mu}^{\left(0\right)}&\stackrel{D\to 4}{\approx}&
\frac{-g^{2}}{8\pi^{2}}\ \gamma_{\mu}\ \frac{1}{\left(4-D\right)}
-\frac{g^{2}}{8\pi^{2}}\ \gamma_{\mu}\ \ln\left(a\mu\right) +
\gamma_{\mu}\ \tilde{{\cal{D}}}
+ \tilde{\cal{C}}_{\mu}\left(\{\epsilon_{\mu}\}, D\right). \nonumber\\
\end{eqnarray}
Hence by combining  (\ref{SB16}) with  (\ref{SB12}) we find again that the terms proportional to $\left(4-D\right)^{-1}$ cancel  (as was the case in the previous two examples), and that we are left with a $\mu$--independent expression with the lattice spacing as the only regulator:
\begin{eqnarray*}
\Lambda_{\mu}\left(p,p',m;a,\{\epsilon_{\mu}\},D=4\right) &=&\frac{-g^{2}}{8\pi^{2}}\gamma_{\mu}\ \ln\left(am\right)+\gamma_{\mu}\tilde{\cal{D}}+
\Lambda_{\mu}^{\left(\mbox{\small{reg.}}\right)}\left(p,p',m\right) \nonumber\\
&&+\sum\limits_{\stackrel{\nu,\tau,\theta=1}{\tau\neq\theta}}^{4}\epsilon_{\theta}\epsilon_{\tau}\gamma_{\mu}\gamma_{\tau}\gamma_{\theta}\ a_{\mu\nu\tau\theta}+\sum\limits_{\stackrel {\tau=1} {\mu\neq\tau} }^{4}\epsilon_{\mu}\epsilon_{\tau}\gamma_{\tau}\ b_{\mu\tau}
+{\cal{O}}\left(a\right)\nonumber\\
\end{eqnarray*}
Here $b_{\mu\tau}\equiv b_{\mu\tau}\left(D=4\right)$ is a finite constant, while $a_{\mu\nu\tau\theta}\equiv a_{\mu\nu\tau\theta}\left(D\to 4\right)=\infty$.
Note that the non--covariant, $\{\epsilon_{\mu}\}$--dependent terms, which destroy the renormalizability of the theory,  vanish after averaging these expressions over all possible sets of $\{\epsilon_{\mu}\}$.
\section{Conclusion}
In this paper we have studied in detail the fermion self energy, vacuum polarization tensor and vertex function for lattice  QED in one--loop order, for the case where the derivative terms in  the Dirac operator are  replaced by one--sided lattice differences. Using the method of ref. \cite{ww} we have systematically expanded the one--loop lattice Feynman integrals for small lattice spacing $a$ up to ${\cal{O}}\left(a^{0}\right)$. Although the small--$a$--expansion method makes use of the dimensional regularization of lattice Feynman integrals in intermediate steps, the in the limit $D\to 4$ divergent terms, possessing a covariant structure,  were found to cancel. While the vacuum polarization tensor was found to have  the correct continuum limit, the fermion self energy and vertex functions included non--covariant terms, which only vanished after averaging the expressions over all possible choices for the one--sided lattice differences. In the case of the vertex function these non--covariant terms were of ${\cal{O}}\left(a^{0}\right)$, while the fermion self energy also included a term of ${\cal{O}}\left(a^{-1}\right)$. By discretizing the derivative in the Dirac operator using a fixed set of one--sided lattice differences, one is led to a non--renormalizable theory. 
\par
Similar computations have been carried out by one of us (N.S.) for the case of Wilson fermions, where it is found that the contribution of ${\cal{O}}\left(a^{-1}\right)$ to the fermion self energy has the structure of a mass term proportional to the Wilson parameter, which can be absorbed into a mass renormalization constant. No non--covariant contributions were encountered in this case.

\bigskip

\par\noindent
{\it Acknowledgment:} We thank  I.O. Stamatescu for valuable comments, and in particular W. Wetzel for several lengthy discussions,  which have been very useful in computing higher order corrections in the small--$a$--expansion.  This work was supported in part by the Graduate College "Systems of many degrees of freedom in physics" of the "Deutsche Forschungsgemeinschaft" (DFG). 
 
\begin{appendix}
\setcounter{section}{1}
\section*{Appendix}\label{appa}
\setcounter{equation}{0}
In all orders of the small--$a$--expansion we have studied, the infrared divergent contributions could be cast into the form
\begin{eqnarray}\label{B1}
M\left(D,n\right) &\equiv&\int\limits_{-\pi}^{+\pi} \frac{d^{D}\hat{k}}{\left(2\pi\right)^{D}} \  \frac{ 1}{ \bigg[ 4\sum\limits_{\rho=1}^{D}\sin^{2}\frac{\hat{k}_{\rho}}{2} \bigg]^{n}}.
\end{eqnarray}
This expression is infrared divergent for $n\geq 2$ and $D\to 4$. The infrared divergent part can be isolated as follows. Define $q^{2}=4\sum\limits_{\rho=1}^{D}\sin^{2}\frac{\hat{k}_{\rho}}{2} $. Then
\begin{eqnarray}\label{B2}
M\left(D,n\right)&=& \left(-1\right)^{n-1}\frac{1}{\left(n-1\right)!} \int\limits_{-\pi}^{+\pi} \frac{d^{D}\hat{k}}{\left(2\pi\right)^{D}} 
\left(\frac{d}{dq^{2}}\right)^{n-1} \int\limits_{0}^{\infty} dt\ e^{-tq^{2}}\nonumber\\
&=&\frac{1}{\left(n-1\right)!} \int\limits_{-\pi}^{+\pi} \frac{d^{D}\hat{k}}{\left(2\pi\right)^{D}} \ \int\limits_{0}^{\infty}dt \ t^{n-1}\  e^{-2t\sum\limits_{\mu=1}^{D} \left(1-\cos\hat{k}_{\mu}\right) }.
\end{eqnarray}
Interchanging the $t$ and $\hat{k}$--integration and setting $s=2t$, one obtains 
\begin{eqnarray}\label{B3}
M\left(D,n\right)&=& \frac{1}{2^{n}\Gamma\left(n\right)}\int\limits_{0}^{\infty}ds\ s^{n-1}\bigg[e^{-s}\ I_{0}\left(s\right)\bigg]^{D},
\end{eqnarray}
where $I_{0}\left(s\right)$ is the modified Bessel function of zero order.  
\par
Consider the case where $n=2$. Then (\ref{B1}) is logarithmically divergent for $D\to 4$. The divergence arises from the large $s$ behaviour  of the integrand in (\ref{B3}) whose leading term, for $n=2$ is given by $\left(2\pi s\right)^{-D/2}$.  We therefore decompose (\ref{B3}) for $n=2$ as follows
\begin{eqnarray}\label{B4}
M\left(D,2\right) &=&\frac{1}{4}\int\limits_{0}^{1}ds\ s\bigg[e^{-s}\ I_{0}\left(s\right)\bigg]^{D} + 
\frac{1}{4}\int\limits_{1}^{\infty}ds\ s\ \Biggl\{\bigg[e^{-s}\ I_{0}\left(s\right)\bigg]^{D}-\frac{1}{\left(2\pi s\right)^\frac{D}{2}}\Biggl\}\nonumber\\
&&+
 \frac{1}{4\left(2\pi\right)^{\frac{D}{2}}}\int\limits_{1}^{\infty}ds\ s^{1-\frac{D}{2}}.
\end{eqnarray}
The first two terms on the rhs are finite, and are denoted below by $\tilde{\cal{N}}_{2}$. For $D>4$ this expression therefore takes the form 
\begin{eqnarray}\label{B5}
M\left(D,2\right)&=&  \tilde{\cal{N}}_{2} -\frac{1}{8\pi^{2}}\   \frac{1}{\left(4-D\right)}.
\end{eqnarray}
Similar expressions can be obtained for $n>2$. Such expressions are required for isolating the infrared divergent parts in higher orders of the small--$a$--expansion. For example the correction of ${\cal{O}}\left(a^{2}\right)$ to the vacuum polarization tensor calculated with fermionic action using one--sided lattice differences for the derivatives,  requires the knowledge of $M\left(D,3\right)$, which can be shown to be given by
\begin{eqnarray}\label{B6}
M\left(D,3\right)&=&   \tilde{\cal{N}}_{3} -\frac{1}{64\pi^{2}}\   \frac{1}{\left(4-D\right)},
\end{eqnarray}
where $\tilde{\cal{N}}_{3}$ is a finite constant for $D\to 4$. 
\end{appendix}
\newpage

\newpage
\epsfxsize12cm                  
\parindent0em                   
\epsfxsize12cm                  
\parindent0em                   

\epsfxsize15cm                  
\parindent0em                   

\end{document}